\documentclass[12pt]{article} \usepackage{epsf}

\newif\ifpdf
\def\Figure#1#2{\centering \leavevmode \ifpdf \pdfimage width #2 #1.png
            \else \tthdump{\epsfclipon \epsfxsize=#2 \epsfbox{#1.ps}}\fi}
\def\Section#1{\section{\sc #1}}
\def\D  {{\sf DUSTY}}
\def\E#1{\hbox{$10^{#1}$}}
\def\eq#1{\begin{equation} #1 \end{equation}}
\def\about  {\hbox{$\sim$}}
\def\gaa    {\mathrel{\hbox{\raise.3ex\hbox{$>$}\llap
                                {\lower.8ex\hbox{$\sim$}}}}}
\def\laa    {\mathrel{\hbox{\raise.3ex\hbox{$<$}\llap
                                {\lower.8ex\hbox{$\sim$}}}}}

\def\la     {\hbox{$\laa$}}
\def\x      {\hbox{$\times$}}
\def\mic    {\hbox{$\mu$m}}
\def\tV     {\hbox{$\tau_V$}}
\def\Mo     {\hbox{$M_{\odot}$}}
\def\Lo     {\hbox{$L_{\odot}$}}
\def\Mdot   {\hbox{$\dot{M}$}}
\def\kms    {\hbox{$\rm km\ s^{-1}$}}
\def\Ivezic {Ivezi\'c}
\def\DS     {\displaystyle}
\def\sub#1{_{\rm #1}}
\def\Frad {\hbox{${\cal F}\sub{rad}$}}
\def\Fgrav{\hbox{${\cal F}\sub{grav}$}}
\def\Te   {\hbox{$T_e$}}
\let\q=\qquad

\def\tthdump#1{#1}      

\begin{document}

\title                  {\sc User Manual for DUSTY}

\author{ \v Zeljko \Ivezic\footnote{Current address:
                Department of Astrophysical Sciences,
                Princeton University, Princeton, NJ 08544},
        Maia Nenkova \& Moshe Elitzur
        \\ \\ Department of Physics and Astronomy
        \\    University of Kentucky, Lexington, KY 40506-0055
        \\[0.5in] October, 1999}
\date{}
\maketitle

\vfil
\begin{abstract}

\D\ solves the problem of radiation transport in a dusty environment.  The code
can handle both spherical and planar geometries.   The user specifies the
properties of the radiation source and dusty region, and the code calculates
the dust temperature distribution and the radiation field in it. The solution
method is based on a self-consistent equation for the radiative energy density,
including dust scattering, absorption and emission, and does not introduce any
approximations. The solution is exact to within the specified numerical
accuracy.

\D\ has built in optical properties for the most common types of astronomical
dust and comes with a library for many other grains. It supports various
analytical forms for the density distribution, and can perform a full dynamical
calculation for radiatively driven winds around AGB stars. The spectral energy
distribution of the source can be specified analytically as either Planckian or
broken power-law. In addition, arbitrary dust optical properties, density
distributions and external radiation can be entered in user supplied files.
Furthermore, the wavelength grid can be modified to accommodate spectral
features.  A single \D\ run can process an unlimited number of models, with
each input set producing a run of optical depths, as specified. The user
controls the detail level of the output, which can include both spectral and
imaging properties as well as other quantities of interest.

\end{abstract}

\newpage
\tthdump{\vglue 1in}

This code is copywrited, 1996--99 by Moshe Elitzur, and may not be copied
without acknowledging its origin. Use of this code is not restricted, provided
that acknowledgement is made in each publication.  The bibliographic reference
to this version of \D\ is \Ivezic, \v Z., Nenkova, M. \& Elitzur, M., 1999,
User Manual for \D, University of Kentucky Internal Report, accessible at {\tt
http://www.pa.uky.edu/$\sim$moshe/dusty}.

\tthdump{\bigskip}

Make sure that you have the current version, with the latest options and
problem fixes, by checking the \D\ Web site. To be automatically notified of
these changes, ask to be placed on the \D\ mailing list by sending an e-mail to
moshe@pa.uky.edu.

\newpage

\tableofcontents \vskip 0.5in

\Section{Introduction} \label{Introduction}

The code \D\ was developed at the University of Kentucky by \v Zeljko \Ivezic,
Maia Nenkova and Moshe Elitzur for a commonly encountered astrophysical
problem: radiation from some source (star, galactic nucleus, etc.) viewed after
processing by a dusty region. The original radiation is scattered, absorbed and
reemitted by the dust, and the emerging processed spectrum often provides the
only available information about the embedded object. \D\ can handle both
planar and centrally-heated spherical density distributions.  The solution is
obtained through an integral equation for the spectral energy density,
introduced in \cite{IE97}. The number of independent input model parameters is
minimized by fully implementing the scaling properties of the radiative
transfer problem, and the spatial temperature profile is found from radiative
equilibrium at every point in the dusty region.

On a Convex Exemplar machine, the solution for spherical geometry is produced
in a minute or less for visual optical depth \tV\ up to $\sim$ 10, increasing
to 5--10 min for \tV\ higher than 100. In extreme cases ($\tV \sim 1000$) the
run time may reach 30 min or more. Run times for the slab case are typically
five times shorter. All run times are approximately twice as long on a 300 MHz
Pentium PC.

The purpose of this manual is to help users get quickly acquainted with the
code. Following a short description of the installation procedure (\S2), the
input and output are described in full for the spherical case in \S3 and \S4.
All changes pertaining to the plane-parallel case are described separately in
\S\ref{slab}.  Finally, \S6 describes user control of \D\ itself.

This new version of \D\ is significantly faster than its previous public
release. Because of the addition of many features, the structure of the input
has changed and old input files will not run on the current version.

\Section{Installation}

The FORTRAN source {\tt dusty.f} along with additional files, including five
sample input files, come in a single compressed file dusty.tar.gz. This file
and its unpacking instructions are available at \D's homepage at {\tt
http://www.pa.uky.edu/$\sim$moshe/dusty}. Alternatively, anonymous ftp
gradj.pa.uky.edu, cd dusty/distribution and retrieve all files and
sub-directories.

\D\ was developed on a Pentium PC and has been run also on a variety of Unix
workstations. It is written in standard FORTRAN 77, and producing the
executable file is rather straightforward.  For example, on a Unix machine

\bigskip

  f77 dusty.f -o dusty

\bigskip\noindent If the compilation is successful you can immediately proceed
to run \D\ without any further action. It should produce the output files {\tt
sphere1.out} and {\tt slab1.out}, printed in appendices \ref{sphere1} and
\ref{slab1}, respectively.  On a 300 MHz Pentium PC with \D\ compiled by Visual
FORTRAN under Windows, these files are produced in just under 2 minutes.
Execution times under Linux are roughly three times longer; the Linux/FORTRAN
implementation on the Digital alpha machine appears to be especially poor, the
execution may be as much as ten times longer. Execution times on SUN
workstations vary greatly with the model: about 1:30 min on an Enterprise 3000,
3 min on SPARC Ultra 1 and 6 min on SPARC20. These run-times should provide an
indication of what to expect on your machine. If \D\ compiles properly but the
execution seems to be going nowhere and the output is not produced in the
expected time, in all likelihood the problem reflects insufficient amount of
machine memory. As a first measure, try to close all programs with heavy demand
on system resources, such as ghostview and Netscape, before running \D.  If
this does not help, the problem may be alleviated by reducing \D's memory
requirements. Section \ref{Memory} describes how to do that.

\Section{Input}

A single \D\ run can process an unlimited number of models.  To accomplish
this, \D's input is always the master input file {\tt dusty.inp}\footnote{{\tt
dusty.inp} must be kept with the \D\ executable file in the same directory.},
which lists the names of the actual input files for all models.  These
filenames must have the form {\tt fname.inp}, where {\tt fname} is arbitrary
and can include a full path, so that a single run may produce output models in
different directories. In {\tt dusty.inp}, each input filename must be listed
on a separate line, with the implied extension {\tt .inp} omitted. Since
FORTRAN requires termination of input records with a carriage return, make sure
you press the ``Enter" key after every filename you enter, especially if it is
in the last line of {\tt dusty.inp}.  Empty lines are ignored, as is all text
following the {\tt `\%'} sign (as in \TeX).  This enables you to enter comments
and conveniently switch on and off the running of any particular model.  The
sample {\tt dusty.inp}, supplied with the program, points to the five actual
input files {\tt sphereN.inp} (N = 1--3) and {\tt slabM.inp} (M = 1, 2). Only
{\tt sphere1} and {\tt slab1} will be executed, since the others are commented
out, providing samples of \D's simplest possible input and output. Once they
have been successfully run you may wish to remove the {\tt `\%'} signs from the
other entries, which demonstrate more elaborate input and output, and check the
running of a full sequence. Your output can be verified against the
corresponding sample output files accessible on \D's homepage.

Each model is characterized by properties of the radiation source and the dusty
region, and \D\ produces a set of up to 999 solutions for all the optical
depths specified in the input.  The output file for {\tt fname.inp} is {\tt
fname.out}, containing a summary of the run and a table of the main output
results. Additional output files containing more detailed tables of radiative
and radial properties may be optionally produced.

The input file has a free format, text and empty lines can be entered
arbitrarily. All lines that start with the {\tt `*'} sign are echoed in the
output, and can be used to print out notes and comments. This option can also
be useful when the program fails for some mysterious reason and you want to
compare its output with an exact copy of the input line as it was read in
before processing by \D. The occurrence of relevant numerical input, which is
entered in standard FORTRAN conventions, is flagged by the equal sign `='. The
only restrictions are that all required input entries must be specified, and in
the correct order; the most likely source of an input error is failure to
comply with these requirements.  Recall, also, that FORTRAN requires a carriage
return termination of the file's last line if it contains relevant input.
Single entries are always preceded by the equal sign, `=', and terminated by a
blank, which can be optionally preceded with a punctuation mark.  For example:
{\tt T = 10,000 K} as well as {\tt Temperature = 1.E4 degrees} and simply {\tt
{} = 10000.00} are all equivalent, legal input entries (note that comma
separations of long numbers are permitted).  Some input is entered as a list,
in which case the first member is preceded by `=' and each subsequent member
must be preceded by a blank (an optional punctuation can be entered before the
blank for additional separation); for example, {\tt Temperatures  = 1E4, 2E4
30,000}. Because of the special role of `=' as a flag for input entry, care
must be taken not to introduce any `=' except when required.  All text
following the {\tt `\%'} sign is ignored (as in \TeX) and this can be used to
comment out material that includes `=' signs.  For example, different options
for the same physical property may require a different number of input entries.
By commenting out with {\tt `\%'}, all options may be retained in the input
file with only the relevant one switched on.

The input contains three types of data --- physical parameters, numerical
accuracy parameters, and flags for optional output files.  The physical
parameters include characteristics of the external radiation, properties of the
dust grains, and the envelope density distribution.  Detailed description of
the program input follows, including examples marked with the `$\bullet$' sign.
Each example contains a brief explanation, followed by sample text typeset in
{\tt typewriter font} as it would appear in the input file. The sample input
files {\tt sphereN.inp} and {\tt slabM.inp}, supplied with \D, are heavily
commented to ease initial use, and can be used as templates.

\subsection
                        {External Radiation}
\label{source}

In the spherical case, \D\ assumes that the external radiation comes from a
point source at the center of the density distribution. Thanks to scale
invariance, the only relevant property of the external radiation under these
circumstances is its spectral shape (see \cite{IE97}).  Six different flag
selected input options are available. The first three involve entry in
analytical form:

\begin{enumerate}

\item

A combination of up to 10 black bodies, each described by a Planck function of
a given temperature. Following the spectrum flag, the number of black bodies is
specified, followed by a list of the temperatures.  When more then one
black-body is specified, the temperature list must be followed by a list of the
fractional contributions of the different components to the total luminosity.

\begin{itemize}
\item A single black body:
\begin{verbatim}
               Spectrum =  1
           Number of BB =  1
            Temperature = 10,000 K
 \end{verbatim}

This could also be entered on a single line as

\hskip 1in {\tt type = 1, N = 1, T = 1E4}

\item Two black bodies, e.g. a binary system, with the first one contributing
80\% of the total luminosity; note that the distance between the stars must be
sufficiently small that the assumption of a central point source remain valid:
\begin{verbatim}
               Spectrum =  1
           Number of BB =  2
           Temperatures = 10,000, 2,500 K
           Luminosities = 4, 1
\end{verbatim}

\end{itemize}

\item
Engelke-Marengo function.  This expression improves upon the black-body
description of cool star emission by incorporating empirical corrections for
the main atmospheric effects. Engelke \cite{Engelk} found that changing the
temperature argument of the Planck function from $T$ to $0.738\,T[1 +
79450/(\lambda T)]^{0.182}$, where $T$ is in K and $\lambda$ is wavelength in
\mic, adequately accounts for the spectral effect of H$^-$. Massimo Marengo
\cite{Mareng} devised an additional empirical correction for molecular SiO
absorption around 8 \mic, and has kindly made his results available to DUSTY.
The selection of this combined Engelke-Marengo function requires as input the
temperature and the relative (to the continuum) SiO absorption depth in~\%.

\begin{itemize}
\item Stellar spectrum parametrized with Engelke--Marengo function:
\begin{verbatim}
                     Spectrum = 2
                  Temperature = 2500 K
         SiO absorption depth = 10 percents
 \end{verbatim}
\end{itemize}

\item
Broken power law of the form:
$$
 \lambda F_\lambda \propto \cases{
        0       &  $\phantom{\lambda(1) < {}} \lambda \le \lambda(1)$   \cr
        \lambda^{-k(1)} &  $\lambda(1) < \lambda \le \lambda(2)$        \cr
        \lambda^{-k(2)} &  $\lambda(2) < \lambda \le \lambda(3)$        \cr
        \vdots                                                          \cr
        \lambda^{-k(N)} &  $\lambda(N) < \lambda \le \lambda(N + 1)$    \cr
                0       &  $\lambda(N+1) < \lambda$                     \cr}
$$

In this case, after the option selection the number $N$ is entered, followed by
a list of the break points $\lambda(i)$, $i = 1\dots N+1$, in \mic\ and a list
of the power indices $k(i)$, $i = 1\dots N$.  The wavelengths $\lambda(i)$ must
be listed in increasing order.

\begin{itemize}
\item A flat spectrum confined to the range 0.1--1.0 \mic:
\begin{verbatim}
               Spectrum = 3
                      N = 1
                 lambda = 0.1, 1 micron
                      k = 0
\end{verbatim}
All spectral points entered outside the range covered by \D's wavelength grid
are ignored. If the input spectrum does not cover the entire wavelength range,
all undefined points are assumed zero.
\end{itemize}
\end{enumerate}

The other three options are for entry in numerical form as a separate
user-supplied input file which lists either (4) $\lambda F_\lambda$ (= $\nu
F_\nu$) or (5) $F_\lambda$ or (6) $F_\nu$ vs $\lambda$.  Here $\lambda$ is
wavelength in \mic\ and $\nu$ the corresponding frequency, and $F_\lambda$ or
$F_\nu$ is the external flux density in arbitrary units.

\begin{enumerate}
\setcounter{enumi}{3} \tthdump{\item}

Stellar spectrum tabulated in a file. The filename for the input spectrum must
be entered separately in the line following the numerical flag. This input file
must have a three-line header of arbitrary text followed by a two-column
tabulation of $\lambda$ and $\lambda F_\lambda$, where $\lambda$ is in \mic\
and $\lambda F_\lambda$ is in arbitrary units. The number of entry data points
is limited to a maximum of 10,000 but is otherwise arbitrary. The tabulation
must be ordered in wavelength but the order can be either ascending or
descending. If the shortest tabulated wavelength is longer than 0.01 \mic, the
external flux is assumed to vanish at all shorter wavelengths.  If the longest
tabulated wavelength is shorter than 3.6 cm, \D\ will extrapolate the rest of
the spectrum with a Rayleigh-Jeans tail.

\begin{itemize}
\item Spectrum tabulated in file {\tt quasar.dat}:

{\tt Spectrum = 4

quasar.dat}
\end{itemize}

\item  Stellar spectrum read from a file as in the previous option, but
$F_\lambda$ is specified (in arbitrary units) instead of $\lambda F_\lambda$.

\begin{itemize}
\item Kurucz model atmosphere tabulated in file {\tt kurucz10.dat}:

{\tt Spectrum = 5

kurucz10.dat}
\end{itemize}

\item  Stellar spectrum read from a file as in the previous option, but
$F_\nu$ is specified (in arbitrary units) instead of $F_\lambda$.
\end{enumerate}

In the last three entry options, the filename for the input spectrum must be
entered separately in the line following the numerical flag. Optionally, you
may separate the flag line and the filename line by an arbitrary number of
lines that are either empty or commented out (starting with {\tt `\%'}). The
files quasar.dat and kurucz10.dat are distributed with DUSTY.

\subsection               {Dust Properties}

Dust optical properties are described by the dust absorption and scattering
cross-sections, which depend on the grain size and  material. Currently, \D\
supports only single-type grains, namely, a single size and chemical
composition.  Grain mixtures can still be treated, simulated by a single-type
grain constructed from an appropriate average.  This approximation will be
removed in future releases which will provide full treatment of grain mixtures.

\subsubsection          {Chemical Composition}
\label{chemistry}

\D\ contains data for the optical properties of six common grain types.  In
models that utilize these standard properties, the only input required is the
fractional abundance of the relevant grains.  In addition, optical properties
for other grains can be supplied by the user.  In this case, the user can either
specify directly the absorption and scattering coefficients or have \D\
calculate them from provided index of refraction. The various alternatives are
selected by a flag, as follows:

\begin{enumerate}

\item \D\ contains data for six common grain types: `warm' and `cold' silicates
from Ossenkopff et al (\cite{Oss92}, {\tt Sil-Ow} and {\tt Sil-Oc}); silicates
and graphite grains from Draine and Lee (\cite{DL84}, {\tt Sil-DL} and {\tt
grf-DL}); amorphous carbon from Hanner (\cite{Hann88}, {\tt amC-Hn}); and SiC
by P\`egouri\`e (\cite{Peg88}, {\tt SiC-Pg}).  Fractional number abundances
must be entered for all these grain types, in the order listed.

\begin{itemize}
\item Mixture containing only dust grains with built-in data for optical
properties:

\begin{verbatim}
   optical properties index = 1
   Abundances for supported grain types, standard ISM mixture:

       Sil-Ow  Sil-Oc  Sil-DL  grf-DL  amC-Hn   SiC-Pg
   x =  0.00    0.00    0.53    0.47    0.00     0.00
 \end{verbatim}
\end{itemize}
The overall abundance normalization is arbitrary.  In this example, the
silicate and graphite abundances could have been entered equivalently as 53 and
47, respectively.

\item With this option, the user can introduce up to ten additional grain types
on top of those built-in.  First, the abundances of the six built-in types of
grains are entered as in the previous option. Next, the number ($\le 10$) of
additional grain types is entered, followed by the names of the data files,
listed separately one per line, that contain the relevant optical properties.
These properties are specified by the index of refraction, and \D\ calculates
the absorption and scattering coefficients using Mie theory.  Each data file
must start with seven header lines (arbitrary text) followed by a three-column
tabulation of wavelength in \mic, and real ({\tt n}) and imaginary ({\tt k})
parts of the index of refraction. The number of table entries is arbitrary, up
to a maximum of 10,000. The tabulation must be ordered in wavelength but the
order can be either ascending or descending. \D\ will linearly interpolate the
data for {\tt n} and {\tt k} to its built-in wavelength grid.  If the supplied
data do not fully cover \D's wavelength range, the refraction index will be
assumed constant in the unspecified range, with a value equal to the
corresponding end point of the user tabulation. The file list should be
followed by a list of abundances, entered in the same order as the names of the
corresponding data files.

\begin{itemize}
\item Draine \& Lee graphite grains with three additional grain types whose {\tt
n} and {\tt k} are provided by the user in data files {\tt amC-zb1.nk}, {\tt
amC-zb2.nk} and {\tt amC-zb3.nk}, distributed with DUSTY.  These files tabulate
the most recent properties for amorphous carbon by Zubko et al \cite{Zubko}:

\begin{verbatim}
   Optical properties index = 2
   Abundances of built-in grain types:
         Sil-Ow  Sil-Oc  Sil-DL  grf-DL amC-Hn SiC-Pg
     x =  0.00    0.00    0.00    0.22   0.00  0.00

   Number of additional components = 3, properties listed in files
                     amC-zb1.nk
                     amC-zb2.nk
                     amC-zb3.nk
   Abundances for these components = 0.45, 0.10, .23
 \end{verbatim}
\end{itemize}

\item This option is similar to the previous one, only the absorption and
scattering coefficients are tabulated instead of the complex index of
refraction, so that the full optical properties are directly specified and there
is no further calculation by \D.  The data filename is listed in the line
following the option flag.  This file must start with a three-line header of
arbitrary text followed by a three-column tabulation of wavelength in \mic,
absorption ($\sigma_{\rm abs}$) and scattering ($\sigma_{\rm sca}$) cross
sections. Units for $\sigma_{\rm abs}$ and $\sigma_{\rm sca}$ are arbitrary,
only
their spectral variation is relevant. The number of entries is arbitrary, with a
maximum of 10,000. The handling of the wavelength grid is the same as in the
previous option.

\begin{itemize}
\item  Absorption and scattering cross sections from the file {\tt
ism-stnd.dat}, supplied with \D, listing the optical properties for the
standard interstellar dust mixture:

{\tt Optical properties index = 3; cross-sections entered in file

\hskip 0.5in        ism-stnd.dat}
\end{itemize}

\end{enumerate}

\D's distribution includes a library of data files with the complex refractive
indices of various compounds of common interest. This library is described in
appendix \ref{nklib}.

\subsubsection          {Grain Size Distribution}

The grain size distribution must be specified only when the previous option was
set to {\tt 1} or {\tt 2}.  When the dust cross sections are read from a file
(previous option set at {\tt 3}), the present option is skipped.

\D\ recognizes two distribution functions for grain sizes $n(a)$ --- the MRN
\cite{MRN77} power-law with sharp boundaries
\eq{
         n(a) \propto a^{-q} \qquad \hbox{for} \quad
                a_{\rm min} \le a \le a_{\rm max}
}
and its modification by Kim, Martin and Hendry \cite{KMH94}, which replaces the
upper cutoff with a smooth exponential falloff
\eq{
  n(a) \propto a^{-q} e^{-a/a_0} \qquad \hbox{for} \quad a \ge a_{\rm min}
}
\D\ contains the standard MRN parameters $q$ = 3.5, $a_{\rm min}$ = 0.005 \mic\
and $a_{\rm max}$ = 0.25 \mic\ as a built-in option.  In addition, the user may
select different cutoffs as well as power index for both distributions.

\begin{enumerate}

\item This is the standard MRN distribution.  No input required other than the
option flag.

\item Modified MRN distribution.  The option flag is followed by listing of the
power index $q$, lower limit $a_{\rm min}$ and upper limit $a_{\rm max}$ in
\mic.

\begin{itemize}
\item Standard MRN distribution can be entered with this option as:

{\tt  Size distribution = 2

 q = 3.5, a(min) = 0.005 micron, a(max) = 0.25 micron}

\item Single size grains with $a$ = 0.05 \mic:

\begin{verbatim}
     Size distribution = 2
                     q = 0 (it is irrelevant in this case)
                a(min) = 0.05 micron
                a(max) = 0.05 micron
     \end{verbatim}

\end{itemize}

\item KMH distribution.  The option flag is followed by a list of the power
index $q$, lower limit $a_{\rm min}$ and the characteristic size $a_0$ in \mic.

\begin{itemize}
\item Size distribution for grains in the dusty envelope around IRC+10216 as
obtained by Jura \cite{Jura} and verified in \Ivezic\ \& Elitzur \cite{IE96b}:

{\tt  Size distribution = 3

 q = 3.5, a(min) = 0.005 micron, a0 = 0.2 micron}

\end{itemize}
\end{enumerate}

\subsubsection{Dust Temperature on Inner Boundary}
\label{Td}

The next input entry is the dust temperature $T_1$ (in K) on the shell inner
boundary.  {\em This is the only dimensional input required by the dust
radiative transfer problem} \cite{IE97}. $T_1$ uniquely determines $F_{e1}$,
the external flux entering the shell, which is listed in \D's output (see \S
\ref{default}). In principle, different types of grains can have different
temperatures at the same location. However, \D\ currently treats mixtures as
single-type grains whose properties average the actual mix. Therefore, only one
temperature is specified.

\subsection{Density Distribution}
\label{density}

In spherical geometry, the density distribution  is specified in terms of the
scaled radius
$$
                        y = {r \over r_1}
$$
where $r_1$ is the shell inner radius.  This quantity is irrelevant to the
radiative transfer problem \cite{IE97}, therefore it is never entered. ($r_1$
scales with the luminosity $L$ as $L^{1/2}$ when all other parameters are held
fixed. The explicit relation is provided as part of \D's output; see \S
\ref{default}.) The density distribution is described by the dimensionless
profile $\eta(y)$, which \D\ normalizes according to $\int\eta dy = 1$. Note
that the shell inner boundary is always $y = 1$.  Its outer boundary in terms
of scaled radii is the shell relative thickness, and is specified as part of
the definition of $\eta$.

\D\ provides three methods for entering the spherical density distribution:
prescribed analytical forms, hydrodynamic calculation of winds driven by
radiation pressure on dust particles, and numerical tabulation in a file.

\subsubsection          {Analytical Profiles}

\D\ can handle three types of analytical profiles: piecewise power-law,
exponential, and an analytic approximation for radiatively driven winds.  The
last option is described in the next subsection on winds.

\begin{enumerate}

\item  Piecewise power law:
$$
 \eta(y) \propto \cases{
        y^{-p(1)}    &  $\phantom{y()}1   \le y < y(1)$       \cr
        y^{-p(2)}    &  $y(1) \le y < y(2)$       \cr
        y^{-p(3)}    &  $y(2) \le y < y(3)$       \cr
                     &  \qquad $\vdots$          \cr
        y^{-p(N)}    &  $y(N - 1) \le y \le y(N)$ \cr}
$$
After the option selection, the number $N$ is entered, followed by a list of the
break points $y(i)$, $i = 1\dots N$, and a list of the power indices $p(i)$, $i
= 1\dots N$.  The list must be ascending in $y$. Examples:

\begin{itemize}

\item Density falling off as $y^{-2}$ in the entire shell, as in a steady-state
wind with constant velocity.  The shell extends to 1000 times its inner radius:

\begin{verbatim}
   density type = 1;     N = 1;   Y = 1.e3;    p = 2
\end{verbatim}

\item Three consecutive shells with density fall-off softening from $y^{-2}$ to
a constant distribution as the radius increases by factor 10:

\begin{verbatim}
              density type = 1
                         N = 3
          transition radii =   10   100    1000
          power indices    =    2     1       0
\end{verbatim}
\end{itemize}

\item   Exponentially decreasing density distribution
\eq{
          \eta \propto  \exp\left(-\sigma\, \frac{y - 1}{Y - 1}\right)
}
where $Y$ is the shell's outer boundary and $\sigma$ determines the fall-off
rate. Following the option flag, the user enters $Y$ and $\sigma$.

\begin{itemize}
\item Exponential fall-off of the density to $e^{-4}$ of its inner value at the
shell's outer boundary $Y = 100$:

\hskip 0.5in {\tt  density type = 2; Y = 100; sigma = 4 }
\end{itemize}
\end{enumerate}

\subsubsection          {Radiatively Driven Winds}
\label{winds}

The density distribution options 3 and 4 are offered for the modeling of
objects such as AGB stars, where the envelope expansion is driven by radiation
pressure on the dust grains. \D\ can compute the wind structure by solving the
hydrodynamics equations, including dust drift and the star's gravitational
attraction, as a set coupled to radiative transfer.  This solution is triggered
with {\tt density type = 3}, while {\tt density type = 4} utilizes an analytic
approximation for the dust density profile which is appropriate in most cases
and offers the advantage of a much shorter run time.

\begin{enumerate}
\setcounter{enumi}{2} \tthdump{\item}
An exact calculation of the density structure from a full dynamics calculations
(see \cite{IE95} and references therein).  The calculation is performed for a
typical wind in which the final expansion velocity exceeds 5 \kms, but is
otherwise arbitrary. The only input parameter that needs to be specified is the
shell thickness $Y = r_{\rm out}/r_1$.

\begin{itemize}
\item
Numerical solution for radiatively driven winds, extending to a distance $10^4$
times the inner radius:

\begin{verbatim}
   density type = 3;     Y = 1.e4
\end{verbatim}
\end{itemize}
The steepness of the density profile near the wind origin increases with
optical depth, and with it the numerical difficulties.  DUSTY handles the full
dynamics calculation for models that have \tV\ \la\ 1,000, corresponding to
\Mdot\ \about\ 4\x\E{-4} \Mo\ $\rm yr^{-1}$.

\item
When the variation of flux-averaged opacity with radial distance is negligible,
the problem can be solved analytically \cite{IEprep}.  In the limit of
negligible drift, the analytic solution takes the form
\eq{
    \eta \propto {1\over y^2}\left[{y \over y - 1 + (v_1/v_e)^2}\right]^{1/2}
}
This density profile provides an excellent approximation under all
circumstances to the actual results of detailed numerical calculations
(previous option). The ratio of initial to final velocity, $\epsilon_v =
v_1/v_e$, is practically irrelevant as long as $\epsilon_v$ \la\ 0.2. The
selection {\tt density type = 4} invokes this analytical solution with the
default value $\epsilon_v = 0.2$. As for the previous option, the only input
parameter that needs to be specified in this case is the outer boundary $Y$.

\begin{itemize}
\item
Analytical approximation for radiatively driven winds, the shell relative
thickness is $Y = 10^4$:

\begin{verbatim}
   density type = 4;     Y = 1.e4
\end{verbatim}
\end{itemize}
Run times for this option are typically 2--3 times shorter and it can handle
larger optical depths than the previous one. Although this option suffices for
the majority of cases of interest, for detailed final fitting you may wish to
switch to the former.

\end{enumerate}

\subsubsection          {Tabulated Profiles}

Arbitrary density profiles can be entered in tabulated form in a file.  The
tabulation could be imported from another dynamical calculation (e.g., star
formation), and \D\ would produce the corresponding IR spectrum.

\begin{enumerate}
\setcounter{enumi}{4} \tthdump{\item}
The input filename must be entered separately in the line following the
numerical flag. This input file must consist of a three-line header of
arbitrary text, followed by a two-column tabulation of radius and density,
ordered in increasing radius.  The inner radius (first entry) corresponds to
the dust temperature $T_1$, entered previously (\S \ref{Td}).  Otherwise, the
units of both radius and density are arbitrary; \D\ will transform both to
dimensionless variables. The number of entry data points is limited to a
maximum of 1,000 but is otherwise arbitrary. \D\ will transform the table to
its own radial grid, with typically \about\ 20--30 points.

\begin{itemize}
\item Density profile tabulated in the file {\tt collapse.dat}:

\begin{verbatim}
   density type = 5;  profile supplied in the file:
                      collapse.dat
\end{verbatim}
\end{itemize}

This file is supplied with \D\ and contains tabulation of the profile $\eta
\propto y^{-3/2}$, corresponding to steady-state accretion to a central mass.

\end{enumerate}

In all cases, care must be taken that $\eta$ not become so small that roundoff
errors cause spline oscillations and decrease accuracy.  To avoid such
problems, \D\ will stop execution with a warning message whenever $\eta$ dips
below \E{-12} or its dynamic range exceeds \E{12}.  This is particularly
pertinent for very steep density profiles, where the outer boundary should be
chosen with care.

\subsection{Optical Depth}

For a given set of the parameters specified above, \D\ will generate up to 999
models with different overall optical depths.  The list of optical depths can
be specified in two different ways.  \D\ can generate a grid of optical depths
spaced either linearly or logarithmically between two end-points specified in
the input.  Alternatively, an arbitrary list can be entered in a file.

\begin{enumerate}

\item Optical depths covering a specified range in linear steps:  Following the
option selection, the fiducial wavelength $\lambda_0$ (in \mic) of optical
depth $\tau_0$ is entered.  The $\tau_0$ grid is then specified by its two ends
and the number of points ($\le 999$).

\begin{itemize}
\item Models with 2.2 \mic\ optical depths including all the integers from 1 to
100:

\begin{verbatim}
       tau grid = 1
       lambda0 = 2.2 micron
       tau(min) = 1; tau(max) = 100
       number of models = 100
\end{verbatim}
\end{itemize}

\item Same as the previous option, only the $\tau_0$ range is covered in
logarithmic steps:

\begin{itemize}
\item Three models with visual optical depth $\tau_V$ =  0.1, 1 and 10:
\begin{verbatim}
       tau grid = 2
       lambda0 = 0.55 micron
       tau(min) = 0.1; tau(max) = 10
       number of models = 3
\end{verbatim}
\end{itemize}

\item
Optical depths list entered in a file: The file name is entered on a single
line after the option selection. The (arbitrary) header text of the supplied
file must end with the fiducial wavelength $\lambda_0$, preceded by the equal
sign, `='. The list of optical depths, one per line up to a maximum of 999
entries, is entered next in arbitrary order.  \D\ will sort and run it in
increasing $\tau_0$.

\begin{itemize}
\item Optical depths from the file {\tt taugrid.txt}, supplied with the \D\
distribution:
\begin{verbatim}
       tau grid = 3; grid supplied in file:
       taugrid.dat
\end{verbatim}
The file {\tt taugrid.dat} is used in the sample input files {\tt slab2.inp}
and {\tt sphere3.inp}.
\end{itemize}
\end{enumerate}

\subsection{Numerical Accuracy and Internal Bounds}
\label{numerics}

The numerical accuracy and convergence of \D's calculations are controlled by
the next input parameter, $q_{\rm acc}$. The accuracy is closely related to the
set of spatial and wavelength grids employed by \D. The wavelength grid can be
modified by users to meet their specific needs (see \S\ref{F-Grid}) and it does
not change during execution. The spatial grids are automatically generated and
refined until the fractional error of flux conservation at every grid point is
less than $q_{\rm acc}$. Whenever \D\ calculates also the density profile
$\eta$, the numerical accuracy of that calculation is also controlled by
$q_{\rm acc}$.

The recommended value is $q_{\rm acc} = 0.05$, entered in all the sample input
files. The accuracy level that can be accomplished is related to the number of
spatial grid points and the model's overall optical depth.  When $\tau_V$ \la\
100, fewer than 30 points will usually produce a flux error of \la\ 1\%\
already in the first iteration. However, as $\tau_V$ increases, the solution
accuracy decreases if the grid is unchanged, and finer grids are required to
maintain a constant level of accuracy.  This is done automatically by \D.  The
maximum number of grid points is bound by \D's array dimensions, which are
controlled by the parameter {\tt npY} whose default value is 40. This internal
limit suffices to ensure convergence at the 5\% level for most models with
$\tau_V$ \la\ 1000.\footnote{Convergence and execution speed can be affected by
the input radiation spectral shape.  A hard spectrum heavily weighed toward
short wavelengths, where the opacity is high, can have an effect similar to
large \tV.} If higher levels of accuracy or larger $\tau_V$ are needed, \D's
internal limits on array sizes must be expanded by increasing {\tt npY}, as
described in \S\ref{Memory}.

\subsection{Output Control}

The final input entries control \D's output. The first is a flag that sets the
level of \D's verbosity during execution.  With {\tt verbose = 1}, \D\ will
output to the screen a minimal progress report of its execution. With {\tt
verbose = 2} you get a more detailed report that allows tracing in case of
execution problems. {\tt verbose = 0} suppresses all messages.  The messages
are printed to the standard output device with the FORTRAN statement {\tt
write(*)}.  If you suspect that your system may not handle this properly,
choose {\tt verbose = 0}.

All other output and its control is explained in the next section. Note again
that this section describes only the output for spherical models. All changes
necessitated by the planar geometry are described separately in \S\ref{Slab
Output}.

\Section{Output}

A typical \D\ run generates an enormous amount of information, and the volume
of output can easily get out of hand. To avoid that, \D's default output is a
single file that lists only minimal information about the run, as described
next. All other output is optional and fully controlled by the user.
\S\ref{Optional Output} describes the optional output and its control.

\subsection{Default Output}
\label{default}

\D\ always produces the output file {\tt fname.out} for each model input {\tt
fname.inp}. In addition to a summary of the input parameters, the default
output file tabulates global properties for each of the optical depths covered
in the run. The table's left column lists the sequential number {\tt \#\#\#} of
the model with the fiducial optical depth {\tt tau0} listed in the next column.
Subsequent columns list quantities calculated by \D\ for that {\tt tau0}:

\begin{list}{$\diamond$}{}
\item
{\tt F1} -- the bolometric flux, in $\rm W\ m^{-2}$, at the inner radius $y =
1$. Only the external source contributes to {\tt F1} since the diffuse flux
vanishes there under the point-source assumption. Note that {\tt F1} is {\em
independent} of overall luminosity, fully determined by the scaled solution
(see \cite{IE97}). The bolometric flux emerging from the spherical distribution
is {\tt F1/$Y^2$}.

Any measure of the shell dimension is irrelevant to the radiative transfer
problem and thus not part of \D's calculations.  Still, the shell size can be
of considerable interest in many applications. For convenience, the next three
output items list different measures of the shell size expressed in terms of
redundant quantities such as the luminosity:

\item
{\tt r1(cm)} -- the shell inner radius where the dust temperature is {\tt T1},
specified in the input (\S \ref{Td}).  This radius scales in proportion to
$L^{1/2}$, where $L$ is the luminosity. The tabulated value corresponds to $L =
\E4\ \Lo$.

\item
{\tt r1/rc} -- where {\tt rc} is the radius of the central source.  This
quantity scales in proportion to $(T_e/T_1)^2$, where $T_e$ is the external
radiation effective temperature.  The listed value is for $T_e = 10,000$ K with
two exceptions: when the spectral shape of the external radiation is the Planck
or Engelke-Marengo function, the arguments of those functions are used for
$T_e$.

\item
{\tt theta1} -- the angular size, in arcsec, of the shell inner diameter. This
angle depends on the observer's position and scales in proportion to $F_{\rm
obs}^{1/2}$, where $F_{\rm obs}$ is the observed bolometric flux.  The
tabulated value corresponds to $F_{\rm obs} = \E{-6}\ \rm W\ m^{-2}$.

\item
{\tt Td(Y)} -- the dust temperature, in K, at the envelope's outer edge.

\item
{\tt err} -- the numerical accuracy, in {\tt \%}, achieved in the run.
Specifically, if $r$ is the ratio of smallest to largest bolometric fluxes in
the shell, after accounting for radial dilution, then the error is $(1 - r)/(1
+ r)$. Errors smaller than 1\% are listed as zero.
\end{list}

When the density distribution is derived from a hydrodynamics calculation for
AGB winds (\S\ref{winds}), three more columns are added to {\tt fname.out}
listing the derived mass-loss rate, terminal outflow velocity and an upper
bound on the stellar mass.  These quantities posses general scaling properties
in terms of the luminosity $L$, gas-to-dust mass ratio $r_{\rm gd}$ and dust
grain bulk density $\rho_s$ \cite{IEprep}.  The tabulations are for $L = \E4\
\Lo$, $r_{\rm gd} = 200$ and $\rho_s = 3\ \rm g~cm^{-3}$, and their scaling
properties are:
\begin{list}{$\diamond$}{}
\item
{\tt Mdot} -- the mass loss rate in \Mo\ $\rm yr^{-1}$, scales in proportion to
$L^{3/4}(r_{\rm gd}\rho_s)^{1/2}$.  This quantity has \about\ 30\% inherent
uncertainty because varying the gravitational correction from 0 up to 50\% has
no discernible effect on the observed spectrum.
\item
{\tt Ve} -- the terminal outflow velocity in \kms, scales in proportion to
$L^{1/4}(r_{\rm gd}\rho_s)^{-1/2}$. The provided solutions apply only if this
velocity exceeds 5 \kms.  {\tt Ve} is subject to the same inherent uncertainty
as {\tt Mdot}.
\item
{\tt M$>$} -- an upper limit in \Mo\ on the stellar mass $M$, scales in
proportion to $L/(r_{\rm gd}\rho_s)$.  The effect of gravity is negligible as
long as $M$ is less than 0.5{\tt *M$>$} and the density profile is then
practically independent of $M$.

\end{list}
There is a slight complication with these tabulations when the dust optical
properties are entered using {\tt optical properties = 3} (\S \ref{chemistry}).
With this option, the scattering and absorption cross sections are entered in a
file, tabulated using arbitrary units since only their spectral shape is
relevant for the solution of the radiative transfer problem. However, the
conversion to mass-loss rate requires also the grain size, and this quantity is
not specified when {\tt optical properties = 3} is used.  \D\ assumes that the
entered values correspond to $\sigma/V$, the cross section per grain volume in
$\mic^{-1}$.  If that is not the case, in the above scaling relations replace
$r_{\rm gd}$ with $r_{\rm gd}V/\sigma$.

Finally, \D\ assumes that the external radiation originates in a central point
source. This assumption can be tested with eqs.\ (27) and (28) of \cite{IE97}
which give expressions for the central source angular size and occultation
effect. From these it follows that the error introduced by the point-source
assumption is no worse than 6\% whenever
\eq{\label{Tmin}
        T_e > 2\times\max[T_1, (F_{e1}/\sigma)^{1/4}].
}
Thanks to scaling, \Te\ need not be specified and is entirely arbitrary as far
as \D\ is concerned. However, compliance with the point-source assumption
implies that the output is meaningful only for sources whose effective
temperature obeys eq.\ \ref{Tmin}. For assistance with this requirement, {\tt
fname.out} lists the lower bound on $T_e$ obtained from this relation for
optically thin sources. Since $F_{e1}$ decreases with optical depth (see
\cite{IE97}), the listed bound ensures compliance for all the models in the
series. However, in optically thick cases $F_{e1}$ may become so small that the
listed bound will greatly exceed the actual limit from eq.\ \ref{Tmin}. In
those cases, the true bounds can be obtained, if desired, from eq.\ \ref{Tmin}
and the model tabulated {\tt F1} (note again that with the point source
assumption, {\tt F1} = $F_{e1}$).

Black-body emission provides an absolute upper bound on the intensity of any
thermal source. Therefore, input radiation whose spectral shape is the Planck
function at temperature $T$ is subject to the limit $T_e \le T$ even though
$T_e$ is arbitrary in principle. In such cases $T$ must comply with eq.\
\ref{Tmin}, otherwise \D's output is suspect and in fact could be meaningless.
\D\ issues a stern warning after the tabulation line of any model with input
spectral shape that is either the Planck or Engelke-Marengo function whose
temperature violates eq.\ \ref{Tmin}.

\subsection{Optional Output}
\label{Optional Output}

In addition to the default output, the user can obtain numerous tabulations of
spectra, imaging profiles and radial distributions of various quantities of
interest for each of the optical depths included in the run. This additional
output is controlled through flags entered at the end of the input file {\tt
fname.inp} that turn on and off the optional tabulations.  Setting all flags to
0, as in {\tt sphere1.inp} and {\tt slab1.inp}, suppresses all optional
tabulations and results in minimal output. A non-zero output flag triggers the
production of corresponding output, occasionally requiring additional input.
Further user control is provided by the value of the output flag. When a
certain flag is set to 1, the corresponding output is listed in a single file
that contains the tabulations for all the optical depth solutions. Setting the
flag to 2 splits the output, when appropriate, tabulating the solution for each
optical depth in its own separate file. This may make it more convenient for
plotting purposes, for example, at the price of many small files.  A few flags
can also be set to 3, splitting the output even further.

Each of the following subsections describes in detail the optional tabulations
triggered by one of the output flags and any additional input it may require.
Appendix \ref{summary} summarizes all the output flags and the corresponding
output files they trigger, and can be used for quick reference.

\subsubsection{Properties of Emerging Spectra}
\label{fname.spp}

Setting the first optional flag to 1 outputs a variety of spectral properties
for all the model solutions to the file {\tt fname.spp}. The tabulation has
four header lines and starts with the model sequential number {\tt \#\#\#}.
The following columns list the corresponding {\tt tau0} and the scaling
parameter $\Psi$ (see \cite{IE97}) for the model. The subsequent columns list
fluxes $f(\lambda) = \lambda F_\lambda/F$, where $F = \int\!F_\lambda
d\lambda$, for various wavelengths of interest:
\begin{list}{$\diamond$}{}
\item {\tt fV} -- relative emerging flux at 0.55 \mic.
\item {\tt fK} -- relative emerging flux at 2.2 \mic.
\item
{\tt f12} -- relative emerging flux at 12 \mic, convolved with the IRAS
filter for this wavelength.
\end{list}
Next are the IRAS colors, defined for wavelengths $\lambda_1$ and $\lambda_2$
in \mic\ as:
\eq{
  [\lambda_2] - [\lambda_1]
  =  \log{\lambda_2 f(\lambda_2) \over \lambda_1 f(\lambda_1)}
  =  \log{F_\nu(\lambda_2) \over F_\nu(\lambda_1)}
}
Columns 5--7 list, in this order, {\tt C21} = $[25] - [12]$, {\tt C31} = $[60]
- [12]$ and {\tt C43} = $[100] - [60]$. They are followed by tabulations of:

\begin{list}{$\diamond$}{}
\item{\tt b8-13} -- the IRAS-defined spectral slope $\beta_{8-13}$ between
8 and 13 \mic:
$$
    \beta_{8-13} = 4.74\log{f(13) \over f(8)} - 1.0
$$
\item{\tt b14-22} -- the IRAS-defined spectral slope $\beta_{14-22}$ between
14 and 22 \mic:
$$
    \beta_{14-22} = 5.09\log{f(22) \over f(14)} - 1.0
$$
\item {\tt B9.8} -- the relative strength of the 9.8 \mic\ feature defined as
$$
    B_{9.8} = \ln{f(9.8) \over f_c(9.8)},
$$
where $f_c(9.8)$ is the continuum-interpolated flux across the feature.
\item
{\tt B11.3} -- the relative strength of the 11.3 \mic\ feature defined as
above for {\tt B9.8}.
\item{\tt R9.8-18} -- the ratio of the fluxes at 9.8 \mic\ and 18 \mic,
$f(9.8)/f(18)$.

\end{list}

\subsubsection  {Detailed spectra for each model}
\label{fname.s}

The next output flag triggers listing of detailed spectra for each model in the
run.  Setting this flag to 1 produces tables for the emerging spectra of all
models in the single output file {\tt fname.stb}.  Setting the flag to 2 places
each table in its own separate file, where file {\tt fname.s\#\#\#} contains
the tabulation for model number {\tt \#\#\#} in the optical depth sequence
listed in the default output file (\S\ref{default}).

In addition to the emerging spectrum, the table for each model lists separately
the contributions of various components to the overall flux, the spectral shape
of the input radiation, and the wavelength dependence of the total optical
depth. The following quantities are tabulated:
\begin{list}{$\diamond$}{}
\item {\tt lambda} -- the wavelength in \mic
\item
{\tt fTot} -- the spectral shape of the total emerging flux $f(\lambda) =
\lambda F_\lambda/\int\!F_\lambda d\lambda$.  Values smaller than \E{-20} are
listed as 0.
\item{\tt xAtt} -- fractional contribution of the attenuated input radiation
to {\tt fTot}
\item{\tt xDs} -- fractional contribution of the scattered radiation to
{\tt fTot}
\item{\tt xDe} -- fractional contribution of the dust emission to {\tt fTot}
\item{\tt fInp} -- the spectral shape of the input (unattenuated) radiation
\item{\tt tauT} -- overall optical depth at wavelength {\tt lambda}
\item{\tt albedo} -- the albedo at wavelength {\tt lambda}

\end{list}

\subsubsection{Images at specified wavelengths}
\label{imaging}

\begin{figure}
\Figure{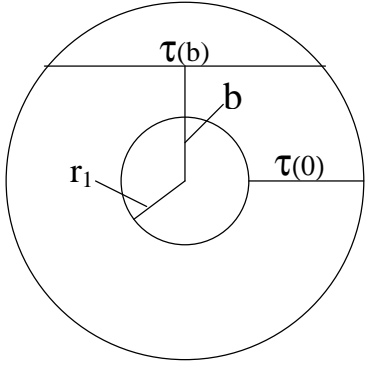}{0.3\hsize}
\caption{Notation for imaging output.}\label{impact parameter}
\end{figure}

The surface brightness is a luminosity-independent self-similar distribution
\cite{IE96a} of $b/r_1$, the impact parameter scaled by the envelope inner
radius (fig. \ref{impact parameter}); note that $r_1$ is listed in the default
output file (\S\ref{default}) for a source luminosity \E4~\Lo. \D\ can produce
maps of the surface brightness at up to 20 wavelengths, specified in the input
file. Setting the option flag to 1 produces imaging tabulations for all the
models of the run in the single output file {\tt fname.itb}, setting the flag
to 2 puts the table for model number {\tt \#\#\#} in its separate file {\tt
fname.i\#\#\#}.

Following the option selection flag, the number ($\le 20$) of desired
wavelengths is entered first, followed by a list of these wavelengths in \mic.

\begin{itemize}
\item Example of additional input data required in {\tt fname.inp} for imaging
output:
\begin{verbatim}
 imaging tables (all models in one file) = 1
 number of wavelengths = 8
 wavelengths = 0.55, 1.0, 2.2, 4, 10, 50, 100, 1000  micron
\end{verbatim}
\end{itemize}
Whenever a specified wavelength is not part of \D's grid, the corresponding
image is obtained by linear interpolation from the neighboring wavelengths in
the grid.  If the nearest wavelengths are not sufficiently close, the
interpolation errors can be substantial. For accurate modeling, all wavelengths
specified for imaging should be part of the grid, modifying it if necessary
(see \S\ref{F-Grid}).

Each map is tabulated with a single header line as follows:
\begin{list}{$\diamond$}{}
\item{\tt b} $= b/r_1$, where $b$ is the impact parameter.
\item
{\tt t(b)} = $\tau(\tt b)/\tau(0)$, where $\tau(\tt b)$ is the overall optical
depth along a path with impact parameter {\tt b}.  Note that $\tau(0)$ is
simply the overall radial optical depth {\tt tauT}, listed in the file {\tt
fname.s\#\#\#} (\S \ref{fname.s}), and that {\tt t(b)} doubles its value across
the shell once the impact parameter exceeds the stellar radius.
\item
The intensity, in Jy arcsec$^{-2}$, at each of the wavelengths listed in the
header line.
\end{list}

A typical image contains a narrow central spike of width $b_c = 2r_c/r_1$,
where $r_c$ is the radius of the central source \cite{IE96a}.  Since this
feature is unresolved in most observations, it is usually of limited interest.
This spike is the only feature of the emerging intensity that depends on the
effective temperature \Te\ of the central source, which is irrelevant to \D's
calculations. The width of the spike scales in proportion to $T_e^{-2}$, its
height in proportion to $T_e^4$. The listed value is for $T_e = 10,000$ K with
two exceptions: when the spectral shape of the external radiation is the Planck
or Engelke-Marengo function, the arguments of those functions are used for
$T_e$.

\subsubsection{Visibilities}
\label{fname.v}

Visibility is the two-dimensional spatial Fourier transform of the surface
brightness distribution (for definition and discussion see \cite{IE96a}). Since
the surface brightness is a self-similar function of $b/r_1$, the visibility is
a self-similar function of $q\theta_1$ where $q$ is the spatial frequency,
$\theta_1 = 2r_1/D$ and $D$ is the distance to the source; note that $\theta_1$
is listed in the default output file for the location where $F_{\rm obs} =
\E{-6}\ \rm W\ m^{-2}$ (\S\ref{default}).

When imaging tables are produced, \D\ can calculate from them the corresponding
visibility functions. The only required input is the flag triggering this
option; if images are not requested in the first place, this entry is skipped.
When the visibility option flag is different from zero, it must be the same as
the one for imaging. Setting both flags to 1 will add visibility tables for all
models to the single file {\tt fname.itb}. Setting the flags to 2 puts the
imaging and visibility tables of each model in the separate file {\tt
fname.i\#\#\#}, setting them to 3 further splits the output by putting each
visibility table in the separate, additional file {\tt fname.v\#\#\#}.

Each visibility table starts with a single header line, which lists the
specified wavelengths in the order they were entered. The first column lists
the dimensionless scaled spatial frequency {\tt q} = $q\theta_1$ and is
followed by the visibility tabulation for the various wavelengths.

\subsubsection{Radial profiles for each model}
\label{fname.r}

The next option flag triggers tabulations of the radial profiles of the
density, optical depth and dust temperature. Setting the flag to 1 produces
tabulations for all the models of the run in the single output file {\tt
fname.rtb}, setting the flag to 2 places the table for model number {\tt
\#\#\#} in its own separate file {\tt fname.r\#\#\#}. The tabulated quantities
are:

\begin{list}{$\diamond$}{}
\item{\tt y} -- dimensionless radius
\item{\tt eta} -- the dimensionless, normalized radial density profile (\S
\ref{density})
\item
{\tt t} -- radial profile of the optical depth variation.  At any wavelength
$\lambda$, the optical depth at radius $y$ measured from the inner boundary is
{\tt t*tauT}, where {\tt tauT} is the overall optical depth at that wavelength,
tabulated in the file {\tt fname.s\#\#\#} (\S \ref{fname.s}).

\item{\tt tauF} -- radial profile of the flux-averaged optical depth
\item
{\tt epsilon} -- the fraction of grain heating due to the contribution of the
envelope to the radiation field (see \cite{IE97}).
\item{\tt Td} -- radial profile of the dust temperature
\item{\tt rg} -- radial profile of the ratio of radiation pressure
to gravitational force, where both forces are per unit volume:
\eq{
    {\Frad\over\Fgrav} = {3L\over16\pi GMc r_{gd}}\,
    {\DS \sum_i n_{d,i} a_i^2\int\!Q_{i,\lambda} f_\lambda\,d\lambda \over
     \DS \sum_i n_{d,i}\rho_{s,i}a_i^3}
}
Here $f_\lambda = F_\lambda/\int F_\lambda d\lambda$ is the local spectral
shape, $\rho_{s,i}$ is the material solid density and $n_{d,i}$ the number
density of grains with size $a_i$.  The gas-to-dust ratio, $r_{gd}$, appears
since the gas is collisionally coupled to the dust. The tabulated value is for
$\rho_s$ = 3 g cm$^{-3}$, $L/M$ = \E4 \Lo/\Mo\ and $r_{gd} = 200$. In the case
of radiatively driven winds $r_{gd}$ varies in the envelope because of the dust
drift, and this effect is properly accounted in the solution. When the dust
optical properties are entered using {\tt optical properties = 3}, grain sizes
are not specified (\S \ref{chemistry}). This case is handled as described in
the last paragraph of \S\ref{default}.

\end{list}

In the case of dynamical calculation with {\tt density type = 3} for AGB stars
(\S\ref{winds}), the following additional profiles are tabulated:

\begin{list}{$\diamond$}{}
\item
{\tt u} -- the dimensionless radial velocity profile normalized to the terminal
velocity {\tt Ve}, which is tabulated for the corresponding overall optical
depth in file {\tt fname.out} (\S \ref{default}).
\item
{\tt drift} -- the radial variation of $v_{\rm g}/v_{\rm d}$, the velocity
ratio of the gas and dust components of the envelope.  This quantity measures
the relative decrease in dust opacity due to dust drift.

\end{list}

\subsubsection{Detailed Run-time messages}
\label{fname.m}

In case of an error, the default output file issues a warning. Optionally,
additional, more detailed run-time error messages can be produced and might
prove useful in tracing the program's progress in case of a failure. Setting
the corresponding flag to 1 produces messages for all the models in the single
output file {\tt fname.mtb}, setting the flag to 2 puts the messages for model
number {\tt \#\#\#} in its own separate file {\tt fname.m\#\#\#}.

\Section{Slab Geometry}
\label{slab}

\D\ offers the option of calculating radiative transfer through a
plane-parallel dusty slab. The slab is always illuminated from the left,
additional illumination from the right is optional.

As long as the surfaces of equal density are parallel to the slab boundaries,
the density profile is irrelevant: location in the slab is uniquely specified
by the optical depth from the left surface.  Unlike the spherical case, there
is no reference to spatial variables since the problem can be solved fully in
optical-depth space.  The other major difference involves the bolometric flux
$F$. In the spherical case the diffuse flux vanishes at $y = 1$ and $F =
F_{e1}/y^2$, where $F_{e1}$ is the external bolometric flux at the shell inner
boundary \cite{IE97}. In contrast, $F$ is constant in the slab and the diffuse
flux does not vanish on either face.  Therefore $F/F_{e1}$, where $F_{e1}$ is
the bolometric flux of the left-side source at slab entry, is another unknown
variable determined by the solution.

The slab geometry is selected by specifying {\tt density type = 0}. The dust
properties are entered as in the spherical case, with the dust temperature
specified on the slab left surface instead of the shell inner boundary. The
range of optical depths, too, is chosen as in the spherical case. The only
changes from the spherical case involve the external radiation and the output.

\begin{figure}
\Figure{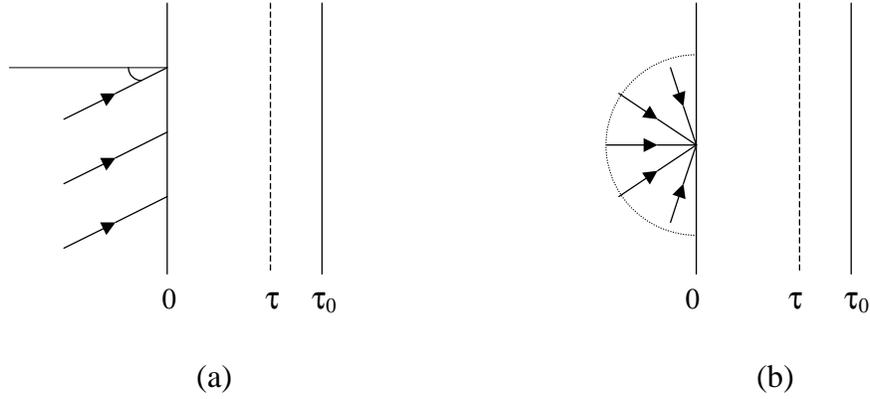}{0.7\hsize}
\caption{The two possible schemes for a slab illuminated from the left.  (a)
Parallel rays impinging at an arbitrary angle to the normal. (b) Isotropic
radiation.}
\end{figure}

\subsection{Illuminating Source(s)}

External radiation is incident from the left side. The presence of an optional
right-side source is specified by a non-zero value for {\tt R}, the ratio of
the right-side bolometric flux at slab entry to that of the left-side source.
Each input radiation is characterized by its spectral shape, which is entered
exactly as in the spherical case (\S\ref{source}), and angular distribution.
The only angular distributions that do not break the planar symmetry involve
parallel rays, falling at some incident angle, and isotropic radiation (see
figure 2). The parallel-rays distribution is specified by the cosine ($> 0.05$)
of the illumination angle, the isotropic distribution is selected by setting
this input parameter to $-1$.  Since oblique angles effectively increase the
slab optical depth, run-times will increase with incidence angle.
\begin{itemize}
\item
Slab geometry with illumination by parallel rays normal to the left surface. In
this case, the spectral shape of the source is entered as in the spherical
case. The only change from the spherical input is that the density profile is
replaced by the following:
\begin{verbatim}
     density type = 0
       cos(angle) = 1.0 (spectral shape entered previously)
                R = 0   (no source on the right)
\end{verbatim}
\end{itemize}
Two-sided illumination is specified by a non-zero {\tt R}, where $0 < {\tt R}
\le 1$. The properties of the right-side source are specified following the
input for {\tt R}.
\begin{itemize}
\item
Slab illuminated from both sides.  The left-side radiation has isotropic
distribution whose spectral shape has been entered previously.  The right-side
source has a bolometric flux half that of the left-side source and a black-body
spectral shape with temperature 3,000 K.  It illuminates the slab with parallel
rays incident at an angle of $60^\circ$ from normal. The density profile is
replaced by the following:
\begin{verbatim}
   density type = 0
     cos(angle) = -1.0 (spectral shape entered previously)
              R = 0.5
   Properties of the right-side source:
     cos(angle) = 0.5
       Spectrum = 1;   N = 1;  Tbb = 3000 K
\end{verbatim}
\end{itemize}

\subsection{Slab Output}
\label{Slab Output}

The output-control flags are identical to those in the spherical case and the
output files are analogous, except for some changes dictated by the different
geometry.

\subsubsection{Default Output}
In the default {\tt fname.out}. the first two columns are the same as in the
spherical case (\S\ref{default}) and are followed by:
\begin{list}{$\diamond$}{}
\item{\tt Fe1} --
bolometric flux, in $\rm W\ m^{-2}$, of the left-side source at the slab left
boundary.
\item{\tt f1} $= F/F_{e1}$, where $F$ is the overall bolometric flux.
Values at and below the internal accuracy of \D's flux computation, \E{-3} when
$q_{\rm acc}$ = 0.05, are listed as zero.
\item{\tt r1(cm)} -- the distance at which a point source with luminosity
\E4 \Lo\ produces the bolometric flux $F_{e1}$.
\item  {\tt Td(K)} -- the dust temperature at the slab right boundary.
\item
{\tt Te(L)} -- the effective temperature, in K, obtained from $F_{e1} =
\sigma{\tt Te}^4$.  When the slab is illuminated also from the right, a column
is added next for {\tt Te(R)}, the effective temperature obtained similarly for
the right-side flux.
\item{\tt err} -- the flux conservation error, defined as in the spherical case.

\end{list}

\subsubsection{Spectral Profiles}
Unlike the spherical case, the slab optional spectral files list properties of
the half-fluxes emerging from both sides of the slab, calculated over the
forward and backward hemispheres perpendicular to the slab faces.  The
magnitudes of the bolometric half-fluxes on the slab right and left faces can
be obtained from tabulated quantities via
 $$ F_{\rm right} = ({\tt R} + {\tt f1})\,{\tt Fe1}, \qquad
    F_{\rm left}  = (1 - {\tt f1})\,{\tt Fe1}. $$
The right-emerging radiation replaces the spherical output in {\tt fname.spp},
{\tt fname.stb} and {\tt fname.s\#\#\#}, analogous tables for the left-emerging
radiation are simply added to the appropriate output files. Setting the
relevant selection flags to 3 places these additional tables in their own
separate files --- {\tt fname.zpp} for spectral properties and {\tt
fname.z\#\#\#} for the detailed spectra of model number {\tt \#\#\#} in the
optical depth sequence.

Similar to the {\tt fTot} column of the spherical case, the spectral shape of
the right-emerging half-flux is printed in column {\tt fRight}.  It consists of
three components whose fractional contributions are listed next, as in the
spherical case: {\tt xAtt} for the left-source attenuated radiation, {\tt xDs}
and {\tt xDe} for the diffuse scattered and emitted components, respectively.
Subsequent columns are as in the spherical case. The tables for the spectral
shape of the left-emerging half-flux {\tt fLeft} are analogous.

\subsubsection{Spatial Profiles}
The output for spatial profiles is similar to the spherical case. The radial
distance {\tt y} and density profile {\tt eta} are removed.  The relative
distance in optical depth from the left boundary, {\tt t}, becomes the running
variable, and the tabulations of {\tt tauF}, {\tt epsilon} and {\tt Td} are the
same (see \ref{fname.r}).  The tabulation for \Frad/\Fgrav\ is dropped,
replaced by three components of the overall bolometric flux: {\tt febol} is the
local net bolometric flux of external radiation; {\tt fRbol} and {\tt fLbol}
are, respectively, the rightward and leftward half-fluxes of the local diffuse
radiation. All components are normalized by {\tt Fe1} so that the flux
conservation relation is {\tt febol + fRbol - fLbol} = {\tt f1} everywhere in
the slab. Note that {\tt fRbol} vanishes on the slab left face, {\tt fLbol} on
the right face.

\bigskip
\D's distribution contains two sample input files, {\tt slab1.inp} and {\tt
slab2.inp}, which can be used as templates for the slab geometry.  The output
generated with {\tt slab1.inp} is shown in appendix \ref{slab1}.

\Section{User Control of \D}

\D\ allows the user control of some of its inner working through tinkering with
actual code statements that control the spatial and spectral grids. The
appropriate statements were placed in the file {\tt userpar.inc} separate from
the main {\tt dusty.f}, and are imbedded during compilation by the FORTRAN
statement {\tt INCLUDE}\footnote{{\tt userpar.inc} must always stay with the
source code in the same directory.}. After modifying statements in {\tt
userpar.inc} , \D\ must be recompiled to enable the changes.

\subsection{Array Sizes for Spatial Grid}
 \label{Memory}

The maximum size of \D's spatial grid is bound by array dimensions. These are
controlled by the parameter {\tt npY} which sets the limit on the number of
radial points.  The default value of 40 must be decreased when \D\ is run on
machines that lack sufficient memory (see \S\ref{Introduction}) and increased
when \D\ fails to achieve the prescribed accuracy (see \S\ref{numerics}). This
parameter is defined in {\tt userpar.inc} via
\begin{verbatim}
      PARAMETER (npY=40)
\end{verbatim}
To modify {\tt npY} simply open {\tt userpar.inc}, change the number 40 to the
desired value, save your change and recompile.  That's all.  Every other
modification follows a similar procedure. Since \D's memory requirements vary
roughly as the second power of {\tt npY}, the maximum value that can be
accommodated on any given machine is determined by the system memory.

The parameter {\tt npY} defines also the size {\tt npP} of the grid used in
angular integrations.  In the case of planar geometry \D\ uses analytic
expressions for these integrations.  Since this grid becomes redundant, {\tt
npP} can be set to unity, allowing a larger maximum {\tt npY}.  The procedure
is described in {\tt userpar.inc}.

\subsection{Wavelength Grid} \label{F-Grid}

\D's wavelength grid is used both in the internal calculations and for the
output of all wavelength dependent quantities. The number of grid points is set
in {\tt userpar.inc} by the parameter {\tt npL}, the grid itself is read from
the file {\tt lambda\_grid.dat}\footnote{{\tt lambda\_grid.dat} must always
stay with the \D\ executable file in the same directory.}. This file starts
with an arbitrary number of text lines, the beginning of the wavelength list is
signaled by an entry for the number of grid points.  This number must be equal
to {\tt npL} entered in {\tt userpar.inc} and to the actual number of entries
in the list.

The grid supplied with \D\ contains 105 points from 0.01 to $3.6\times10^{4}$
\mic.  The short wavelength boundary is to ensure adequate coverage of input
radiation from an O star, for example, which peaks at 0.1 \mic.  Potential
effects on the grain material by such hard radiation are not included in \D.
The long wavelength end is to ensure adequate coverage at all wavelengths where
dust emission is potentially significant. Wavelengths can be added and removed
provided the following rules are obeyed:
\begin{enumerate}
\item
Wavelengths are specified in \mic.
\item
The shortest wavelength must be $\le 0.01$ \mic, the longest $\ge
3.6\times10^{4}$ \mic.
\item
The ratio of each consecutive pair must be $\le$ 1.5.
\end{enumerate}
The order of entries is arbitrary, \D\ sorts them in increasing wavelength and
the sorted list is used for all internal calculations and output.  This
provides a simple, convenient method for increasing the resolution at selected
spectral regions: just add points at the end of the supplied grid until the
desired resolution is attained.  Make sure you update both entries of {\tt npL}
and recompile \D.

In practice, tinkering with the wavelength grid should be reserved for adding
spectral features. Specifying the optical properties of the grains at a
resolution coarser than that of the wavelength grid defeats the purpose of
adding grid points. The optical properties of grains supported by \D\ are
listed on the default wavelength grid.  Therefore, modeling of very narrow
features requires both the entry of a finer grid in {\tt lambda\_grid.dat} and
the input of user-supplied optical properties (see \S\ref{chemistry}) defined
on that same grid.

\vfil

\newpage
\appendix
\section*{\sc Appendices}
 \addtocontents{toc}{\break \vfil}
 \addcontentsline{toc}{section}{\sc Appendices}

\Section{Output Summary} \label{summary}

\D's default output is the file {\tt fname.out}, described in \S\ref{default}.
Additional output is optionally produced through selection flags, summarized in
the following table.  The second column lists the section number where a
detailed description of the corresponding output is provided.

\begin{table}[htbp]
\begin{center}
\renewcommand{\arraystretch}{1.3}

\caption{\hfil Summary of all Output Options}\label{Options Table}
\centerline{}
\renewcommand{\arraystretch}{1.3}
\begin{tabular}{|l|r||c|c|c|}                              \hline
 \multicolumn{1}{|c|}{Output Listing}  &
 \multicolumn{1}{c||}{\S}     &
 \multicolumn{3}{|c|}{Output File Triggered by Flag}  \\ \cline{3-5}
                  & & 1 & 2 & 3 \\ \hline
Spectral properties, all models  & \ref{fname.spp}
                                 & {\tt fname.spp}
                                 & {\tt fname.spp}
                                 & {\tt fname.spp}
                                 \\ \cline{1-2} \cline{5-5}
\q Slab, left-face spectra       & \ref{Slab Output}
                                 & & & {\tt fname.zpp}
                                 \\ \hline
Detailed spectra, each model     & \ref{fname.s}
                                 & {\tt fname.stb}
                                 & {\tt fname.s\#\#\#}
                                 & {\tt fname.s\#\#\#}
                                 \\ \cline{1-2} \cline{5-5}
\q Slab, left-face spectra       & \ref{Slab Output}
                                 & & & {\tt fname.z\#\#\#}
                                 \\ \hline
Images                           & \ref{imaging}
                                 & {\tt fname.itb}
                                 & {\tt fname.i\#\#\#}
                                 & {\tt fname.i\#\#\#}
                                 \\ \cline{1-2} \cline{5-5}
\q Visibilities                  & \ref{fname.v}
                                 &&& {\tt fname.v\#\#\#}  \\
                                 \hline
Radial profiles                  & \ref{fname.r}
                                 & {\tt fname.rtb}
                                 & {\tt fname.r\#\#\#}   &
                                 \\ \hline
Error messages                   & \ref{fname.m}
                                 & {\tt fname.mtb}
                                 & {\tt fname.m\#\#\#}   &
                                 \\ \hline
\end{tabular}
\end{center}
\end{table}

\Section{Pitfalls, Real and Imaginary} \label{pitfalls}

This appendix provides a central depository of potential programming and
numerical problems. Some were already mentioned in the text and are repeated
here for completeness.

\begin{itemize}
\item
FORTRAN requires termination of input records with a carriage return. Make sure
you press the ``Enter" key whenever you enter a filename in the last line of
{\tt dusty.inp}.

\item
In preparing input files, the following two rules must be carefully observed:
(1) all required input entries must be specified, and in the correct order; (2)
the equal sign, `=', must be entered only as a flag to numerical input. When
either rule is violated and \D\ reaches the end of the input file while looking
for additional input, you will obtain the error message:
\begin{verbatim}
     ****TERMINATED. EOF reached by RDINP while looking for input.
     *** Last line read:
\end{verbatim}
This message is a clear sign that the input is out of order.

\item
Linux apparently makes heavier demand on machine resources than Windows.  On
any particular PC and a given value of {\tt npY}, \D\ may execute properly
under Windows but not under Linux, dictating a smaller {\tt npY}.

\item
\D's execution under the Solaris operating system occasionally gives the
following warning message:
\begin{verbatim}
       Note: IEEE floating-point exception flags raised:
            Inexact;  Underflow;
        See the Numerical Computation Guide, ieee_flags(3M)
\end{verbatim}
This ominous message is triggered on Solaris also by other applications and is
not unique to \D. The reason for it is not yet clear and it is not issued on
other platforms. In spite of this statement, the code performs fine and
produces results identical to those on machines that do not issue this warning.

\item
CRAY J90 machines have specific requirements on FORTRAN programs which prevent
\D\ from running in its present form.  If you plan to run \D\ on this platform
you'll have to introduce some changes in the source code, such as replacing all
{\tt DOUBLE PRECISION} statements with {\tt REAL*4} .

\end{itemize}

\vspace {1cm}

\Section{Sample Output File: {\tt sphere1.out}} \label{sphere1}

\begin{verbatim}
 ===========================
  Output from program Dusty
  Version: 2.0
 ===========================

  INPUT parameters from file:
  sphere1.inp

* ----------------------------------------------------------------------
* NOTES:
* This is a simple version of an input file producing a minimal output.
* ----------------------------------------------------------------------
  Central source spectrum described by a black body
  with temperature: 2500 K
  --------------------------------------------
  Abundances for supported grains:
  Sil-Ow Sil-Oc Sil-DL grf-DL amC-Hn SiC-Pg
  1.000  0.000  0.000  0.000  0.000  0.000
  MRN size distribution:
       Power q:  3.5
  Minimal size: 5.00E-03 microns
  Maximal size: 2.50E-01 microns
  --------------------------------------------
  Dust temperature on the inner boundary: 800  K
  --------------------------------------------
  Density described by 1/r**k with k =  2.0
  Relative thickness: 1.000E+03
  --------------------------------------------
  Optical depth at 5.5E-01 microns: 1.00E+00
  Required accuracy: 5%
  --------------------------------------------

  ====================================================
  For compliance with the point-source assumption, the
  following results should only be applied to sources
  whose effective temperature exceeds 1737 K.
  ====================================================

  RESULTS:
  --------
 ###   tau0   F1(W/m2)  r1(cm)    r1/rc   theta1  Td(Y) err
 ###     1        2        3        4        5      6    7
 ==========================================================
   1 1.00E+00 2.88E+04 3.26E+14 8.78E+00 2.43E+00   44   0
 ==========================================================
   (1) Optical depth at 5.5E-01 microns
   (2) Bolometric flux at the inner radius
   (3) Inner radius for L=1E4 Lsun
   (4) Ratio of the inner to the stellar radius
   (5) Angular size (in arcsec) when Fbol=1E-6 W/m2
   (6) Dust temperature at the outer edge (in K)
   (7) Maximum error in flux conservation (%)
 =================================================
  Everything is OK for all models
 ========== THE END ==============================
\end{verbatim}

\Section{Sample Output File: \tt slab1.out} \label{slab1}
\begin{verbatim}
 ===========================
  Output from program Dusty
  Version: 2.0
 ===========================

  INPUT parameters from file:
  slab1.inp

* ----------------------------------------------------------------------
* NOTES:
* This is a simple version of an input file for calculation in
* planar geometry with single source illumination.
* ----------------------------------------------------------------------
  Left-side source spectrum described by a black body
  with temperature: 2500 K
  --------------------------------------------
  Abundances for supported grains:
  Sil-Ow Sil-Oc Sil-DL grf-DL amC-Hn SiC-Pg
  1.000  0.000  0.000  0.000  0.000  0.000
  MRN size distribution:
       Power q:  3.5
  Minimal size: 5.00E-03 microns
  Maximal size: 2.50E-01 microns
  --------------------------------------------
  Dust temperature on the slab left boundary: 800  K
  --------------------------------------------
  Calculation in planar geometry:
  cos of left illumination angle =   1.000E+00
  R =   0.000E+00
  --------------------------------------------
  Optical depth at 5.5E-01 microns: 1.00E+00
  Required accuracy: 5%
  --------------------------------------------


  RESULTS:
  --------
 ###   tau0    Fe1(W/m2)   f1     r1(cm)  Td(K)  Te(L)  err
 ###     1        2         3       4       5      6     7
 ==========================================================
   1 1.00E+00 2.59E+04  9.33E-01 3.43E+14  755 8.22E+02  0
 ==========================================================
   (1) Optical depth at 5.5E-01 microns
   (2) Bol.flux of the left-side source at the slab left boundary
   (3) f1=F/Fe1, where F is the overall bol.flux in the slab
   (4) Position of the left slab boundary for L=1E4 Lsun
   (5) Dust temperature at the right slab face
   (6) Effective temperature of the left source (in K)
   (7) Maximum error in flux conservation (%)
 =================================================
  Everything is OK for all models
 ========== THE END ==============================
\end{verbatim}

\newpage
\Section{Library of Optical Constants} \label{nklib}

\D's distribution includes a library of data files with the complex refractive
indices of various compounds of interest.  The files are standardized in the
format \D\ accepts. Included are the optical constants for the seven built-in
dust types as well as other frequently encountered astronomical dust
components.  This library will be updated continuously at the \D\ site. The
following table lists all the files currently supplied.

\begin{table}[h]
\begin{center}

\caption{\hfil Optical Constants Library Supplied with Dusty} \centerline{}

\begin{tabular}{llrr}     \hline \hline
 \tthdump{\noalign{\medskip}}
 \multicolumn{1}{c}{File Name}    &
 \multicolumn{1}{c}{Compound}     &
 \multicolumn{1}{c}{Range (\mic)} &
 \multicolumn{1}{c}{Ref}
 \\ \tthdump{\noalign{\medskip}}
    \hline
    \tthdump{\noalign{\smallskip}}

{\tt Al2O3-comp.nk} & Al$_2$O$_3$-compact        & 7.8 -- 200   & \cite{Jena}  \\
{\tt Al2O3-por.nk}  & Al$_2$O$_3$-porous         & 7.8 -- 500   & \cite{Jena}  \\
{\tt amC-hann.nk}   & amorphous carbon           & 0.04 -- 905  & \cite{Hann88}\\
{\tt amC-zb1.nk}    & amorphous carbon (BE)      & 0.05 -- 1984 & \cite{Zubko} \\
{\tt amC-zb2.nk}    & amorphous carbon (ACAR)\q  & 0.04 -- 1984 & \cite{Zubko} \\
{\tt amC-zb3.nk}    & amorphous carbon (ACH2)    & 0.04 -- 948  & \cite{Zubko} \\
{\tt crbr300.nk}    & crystalline bronzite       & 6.7 -- 487.4 & \cite{Henn97}\\
{\tt crMgFeSil.nk}  & crystalline silicate       & 6.7 -- 584.9 & \cite{Jena}  \\
{\tt FeO.nk}        & FeO (5.7g/ccm)             & 0.2 -- 500   & \cite{Jena}  \\
{\tt gloliMg50.nk}  & glassy olivine             & 0.2 -- 500   & \cite{Dorsch}\\
{\tt glpyr300.nk}   & glassy pyroxene at 300 K   & 6.7 -- 487   & \cite{Henn97}\\
{\tt glpyrMg50.nk}  & glassy pyroxene            & 0.2 -- 500   & \cite{Dorsch}\\
{\tt glSil.nk}      & glassy silicate            & 0.4 -- 500   & \cite{Jaeger}\\
{\tt grph1-dl.nk}   & graphite, $E \perp c$      & 0.001 -- \E3 & \cite{DL84}  \\
{\tt grph2-dl.nk}   & graphite, $E \parallel c$  & 0.001 -- \E3 & \cite{DL84}  \\
{\tt opyr-pwd.nk}   & ortho-pyroxenes - powder   & 5.0 -- 25    & \cite{Roush} \\
{\tt opyr-slb.nk}   & ortho-pyroxenes - slab     & 5.0 -- 25    & \cite{Roush} \\
{\tt OssOdef.nk}    & O-deficient CS silicate    & 0.4 -- \E4   & \cite{Oss92} \\
{\tt OssOrich.nk}   & O-rich IS silicate         & 0.4 -- \E4   & \cite{Oss92} \\
{\tt SiC-peg.nk}    & $\alpha$-SiC               & 0.03 -- 2000 & \cite{Peg88} \\
{\tt Sil-dlee.nk}   & ``Astronomical silicate"   & 0.03 -- 2000 & \cite{DL84}  \\
{\tt Sil-oss1.nk}   & warm O-deficient silicates & 0.4 -- \E4   & \cite{Oss92} \\
{\tt Sil-oss2.nk}   & cold O-rich silicate       & 0.4 -- \E4   & \cite{Oss92} \\

\noalign{\medskip} \hline
\end{tabular}
\end{center}
\end{table}

\end{document}